\documentclass[epj,final]{svjour}
\usepackage[dvipsnames]{xcolor}
\usepackage{graphicx}
\usepackage{float}
\usepackage{textcomp}
\usepackage{latexsym}
\usepackage{url}
\usepackage{amsfonts}
\usepackage{amsmath, amssymb}
\RequirePackage{graphicx}
\usepackage{caption,subcaption}
\usepackage{indentfirst}

\begin{document}
	\title{Neural Network Analysis of S-Star Dynamics: Implications for Modified Gravity}
        \titlerunning{Neural Network Analysis of S-Star Dynamics}
	\author{N.Galikyan\inst{1,2}, Sh.Khlghatyan\inst{2}, A.A.Kocharyan\inst{3}, V.G.Gurzadyan\inst{2,4,5}
	}                     
	%
	%
	\institute{National Research Nuclear University MEPhI, Moscow, Russia \and Center for Cosmology and Astrophysics, Alikhanian National Laboratory and Yerevan State University, Yerevan, Armenia  \and School of Physics and Astronomy, Monash University, Clayton, Australia \and SIA, Sapienza Universita di Roma, Rome, Italy \and E-mail:gurzadyan@yerphi.am}
	\date{Received: date / Revised version: date}
	%

	\abstract{We studied the dynamics of S-stars in the Galactic center using the physics-informed neural networks. The neural networks are considered for both, Keplerian and the General Relativity dynamics, the orbital parameters for stars S1, S2, S9, S13, S31, and S54 are obtained and the regression problem is solved. It is shown that the neural network is able to detect the Schwarzschild precession for S2 star, while the regressed part revealed an additional precession. Attributing the latter to a possible contribution of a modified gravity, we obtain a constraint for the weak-field modified General Relativity involving the cosmological constant which also deals with the Hubble tension.  Our analysis shows the efficiency of neural networks in revealing the S-star dynamics and the prospects upon the increase of the amount and the accuracy of the observational data.   
}

%
\maketitle

\section{Introduction}

The motion of S-stars in the Galactic center has become an important source for probing the value of the mass of the supermassive black hole of Sgr A* and a natural laboratory for testing of General Relativity (GR) and different theories of gravity. The S-star dynamics is being studied within dedicated observational surveys involving different methods of data analysis,  \cite{GRAVITY} and references therein. An essential area of the observational data analysis data is dedicated to the testing of the modified gravity theories based on the reconstruction of the dynamics of individual S-stars, e.g. \cite{Cap,Fermionic,Borka,Z1,Z2,Yukawa,EinMax,GRAVITYscalarclouds}, thus complementing the recent tests of GR, e.g. \cite{LIGO,Ciu1,Ciu2,Ciu3,Event}. 

In this paper we involve neural networks, namely, the physics-informed neural networks (PINN) \cite{PINN1,PINN2} to analyse the dynamics of the S-stars. 
Neural networks of various architecture are already widely used in broad range of physical problems, from particle physics to astrophysics, e.g. \cite{MLParticle,AIFeynamn,NNOrbMech,NNSubhalo}.

In our analysis we used the neural networks considering both the Newtonian theory and the General Relativity, to explicitly reveal the differences which are able to trace the network architectures. We obtain the orbital parameters of certain stars, which then enables us to constrain a weak-field modified General Relativity. 
As is known, in contrast to the Keplerian orbital motion, in GR there occurs an apsidal precession (for central body without a rotational momentum, i.e. for Schwarzschild metric), yielding during one period of revolution a precession shift \cite{MTW}
\begin{equation}\label{eq:SchPrec}
	\delta\varphi_{\text{SP}}=3\frac{r_g}{a(1-e^2)}\pi,
\end{equation}
where $r_g$ is the gravitational radius of the central body, $a$ is the semi-major axis and $e$ is the eccentricity of the orbit. Since for the S2 star the observational data are available for more than one revolution period, it was involved to test the GR and constrain possible deviations from it \cite{GRAVITY}. In this paper, as an involved additional deviation, we consider the possible contribution to the precession rate by the weak-field modified GR involving the cosmological constant $\Lambda$.  That modification is based on the condition of identity of sphere's and point mass's gravity, and $\Lambda$-term naturally enters as an additional one in the Newtonian force  \cite{G}
\begin{equation}\label{eq:ModF}
\mathbf{F}(r) = \left(-\frac{A}{r^2} + \Lambda r\right)\hat{\mathbf{r}}\ .
\end{equation} 
Then the metric tensor components have the form
\begin{equation} 
g_{00} = 1 - \frac{2 G M}{r c^2} - \frac{\Lambda r^2}{3};\,\,\, g_{rr} = \left(1 - \frac{2 G M}{r c^2} - \frac{\Lambda r^2}{3}\right)^{-1},
\end{equation} 
where the currently estimated value for the cosmological constant is $\Lambda = 1.11\times 10^{-52}\, [\text{m}]^{-2}$ \cite{L}.
This metric is known as Schwarzschild - de Sitter metric \cite{Rind}, and considered from Eq.(1) as weak-field GR, 
it provides a description of astrophysical structures such as the galaxy groups and clusters \cite{GS1,G1,GS5}.

Eq.(1) enables to include the cosmological constant in the McCrea-Milne cosmology \cite{MM,MM1}. The additional $\Lambda$-term will create a non-force-free field inside the spherical shell, and it appears efficient in describing the properties of galactic halos, other observable effects  \cite{SK,SKG1,SKG2}. Its role in the relative instability of $N$-body gravitating systems has been analyzed in \cite{GKS}. The consideration of $\Lambda$ as a fundamental physical constant links the cosmological evolution with the notion of information \cite{GS3}. This approach also enables to describe the Hubble tension as a result of local and global flows \cite{GS4}, to address the structure formation in the local Universe \cite{GFC1,GFC2}.
 
Thus, our first goal is to apply PINN in obtaining the orbital parameters  of S-stars i.e. the eccentricity $e$ and focal parameter $p$, the mass of the central body and solve the regression problem $u(\varphi)$, where $u$ is the inverse of the radius and $(r,\varphi)$ are the polar coordinates of the motion. Then, for the Schwarzschild metric, from the S2 star data we get a constraint for the GR modification involving the cosmological constant. Note that, we do not take into account the interaction between individual stars, i.e. we do not consider the $N$-body problem. The possible role of extended mass distribution on the dynamics of S-stars has been considered earlier (e.g. \cite{Rub,Zakh}), however, the analysis by the Gravity collaboration \cite{Abut} for plausible density profiles shows that the extended mass component inside the S2 star apocenter  must be less than 0.1\% of the mass of the central black hole. As mentioned, we used the data for S2. Future more accurate observations will enable to get more information on the role of extended mass distribution. 

\section{Observational Data}

To analyze the motion of S-stars we used the available observational data as studied in \cite{Gillessen}.
The coordinates of the S-stars motion are given by right ascension and declination, using the Thiele-Innes constants and real orbits argument of periapsis, the inclination, and the ascending node angles we can transform the Sky-plane coordinates into the Cartesian coordinates $(x\,\,[\text{Au}],y\,\,[\text{Au}])_{\text{Star}_{i}}$ \cite{GRAVITY,Fermionic}. Then, we normalize the Cartesian coordinates in two ways:\\
1) Each star data normalized individually $\left(\tilde{x}:=\frac{x}{S_i}, \tilde{y}:=\frac{y}{S_i}\right)_{\text{Star}_i}$, with a normalization coefficient $S_i$ corresponding to the $i$-th star.\\
2) All stars data normalized by a single normalization coefficient $S$ $\left(\tilde{x}:=\frac{x}{S}, \tilde{y}:=\frac{y}{S}\right)_{\text{Star}_i}$.\\
Numerical simulations have demonstrated that the outcomes remain consistent regardless of the specific choice.

\begin{figure}[h!]
    \centering
    \includegraphics[width=0.6\linewidth]{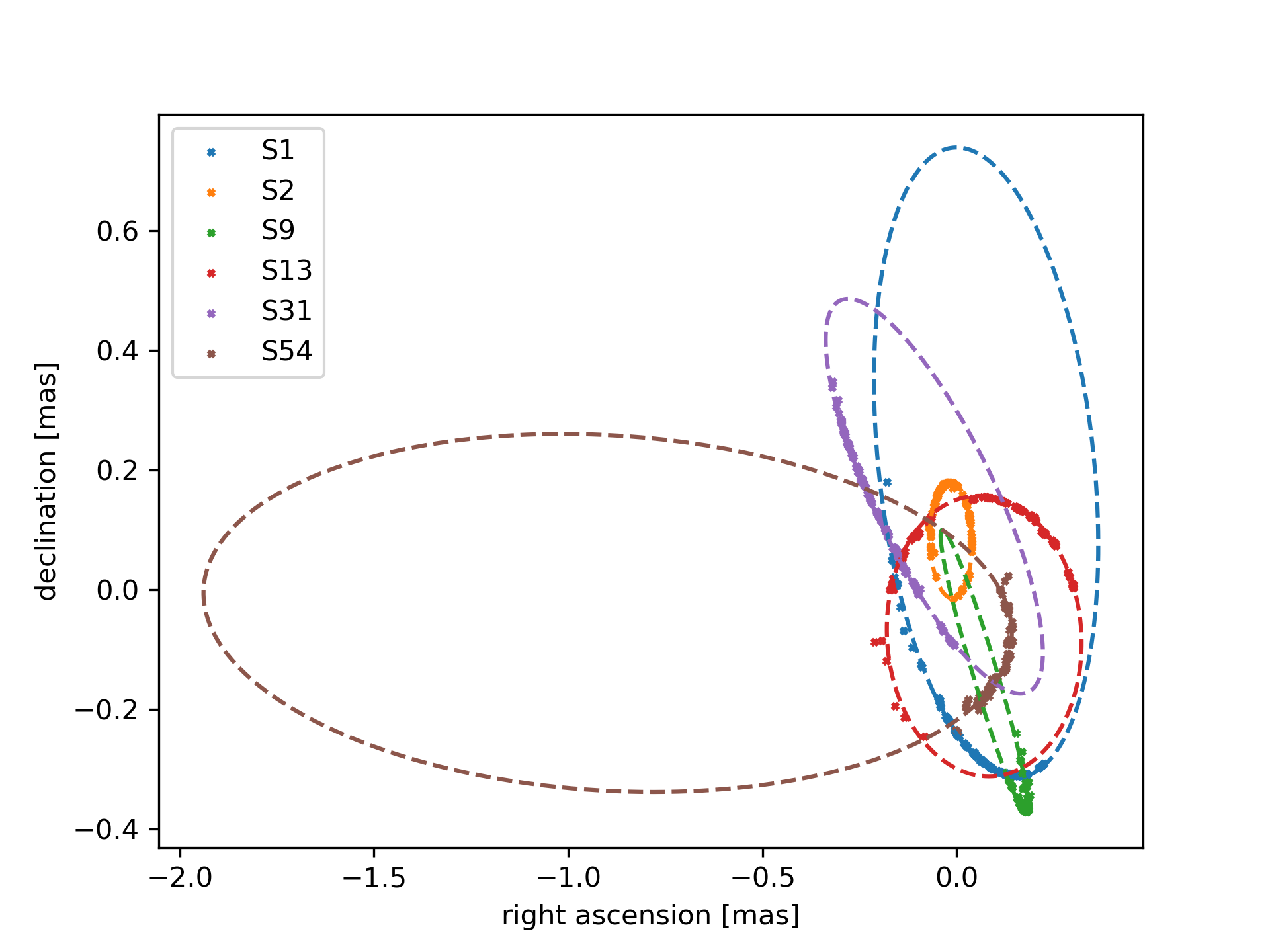}
    \caption{The coordinates of the considered stars: points indicate the observed data and dashed lines are the ellipses obtained from the orbital parameters of the Keplerian fit \cite{Gillessen}.}
    \label{fig:Data}
\end{figure}

We considered the stars given in the Tab.(\ref{tab:data}) and in the Fig.(\ref{fig:Data}) \cite{Gillessen}.
\begin{table}[h!]
    \centering
    \begin{tabular}{|c|c|c|c|c|}
    \hline
    Name    & eccentricity $e$ & focal parameter\ $p\,\,[\text{Au}]$& Period $T\,\,[\text{Yr}]$ & Data count\\
    \hline
     S1    & $0.556$ & $3412$ &  $166$ & $161$\\
     \hline
     S2   & $0.884$ & $228$ & $16$ & $145$\\
     \hline
     S9   & $0.644$ & $1323$ & $51$ & $160$\\
     \hline
     S13  & $0.425$ & $1796$ & $49$ & $127$\\
     \hline
     S31  & $0.550$ & $2601$ & $108$ & $51$\\
     \hline
     S54  & $0.893$ & $2017$ & $477$ & $94$\\
     \hline
    \end{tabular}
    \caption{S-stars the parameters of which were used during the training. The values of orbital parameters and periods were obtained by Keplerian fit \cite{Gillessen}. }
    \label{tab:data}
\end{table}

\section{Neural networks}

\subsection{Implementation of the PINN}

PINNs have a broad range of applications \cite{PINNSurvey}, since they allow combining neural network approaches with physical models, which are often presented in the form of differential equations, both ODE and PDE \cite{PINN1,PINN2}. These neural networks are already used in such areas as the fluid mechanics \cite{Fluid}, nonlinear optics (for solving the nonlinear Schroedinger equation)\cite{NonlinSchrod}, heat transfer problems\cite{Heat}, industrial problems (power systems \cite{Power} and main bearing fatigue prognosis \cite{Fatigue}), medicine (cardiac activation mapping \cite{Cardiac} and cardiovascular flows modeling \cite{Cardio}), etc.

There are different approaches and implementations of PINN, and in this paper we involve PINN to solve regression problem using the differential equations of the corresponding physical models. This approach is based on the so-called physical loss function during its training together with a classic regression loss $L_{reg}(y, f(x))$ (e.g. mean squared error), where $y$ is the ground truth value and $f(x)$ is the model. If the physical process is described by~Eq.(\ref{eq:diff_eq})
\begin{equation} \label{eq:diff_eq}
    F(x, y, y'_x, y_x'', \ldots, y_x^{(n)}) = 0.
\end{equation}
Then the physical loss is given by the following loss function Eq.(\ref{eq:phys_loss})
\begin{equation} \label{eq:phys_loss}
    L_{phys}(f(x), x) = F^2\left(x, f(x), f'(x), f''(x), \ldots, f^{(n)}(x)\right).
\end{equation}
To calculate the total loss value one should use the actual data $\{(y_i, x_i)\}_{i=1}^N$ and sample $\{(\hat{x}_i)\}_{i=1}^{N_p}$ points from a larger data domain. Then the loss function may be calculated by Eq.(\ref{eq:pinn_loss}), where $\alpha$ is a given training step-dependent regularization parameter. 
\begin{equation} \label{eq:pinn_loss}
    \mathcal{L}(f(\cdot), y, x, \hat{x}) = \frac{1}{N}\sum_{i=1}^N L_{reg}(y_i, f(x_i)) + \frac{\alpha}{N_p}\sum_{i=1}^{N_p} L_{phys}(f(\hat{x}_i), \hat{x}_i).
\end{equation}
We involve PINN for the following reasons:
\begin{enumerate}
    \item It can be used in cases when not much data are available;
    \item One can extrapolate the results of the regression to a bigger data domain;
    \item One can consider parameters in the differential equation as trainable parameters for NN, and estimate their values by training.
\end{enumerate}

\subsection{Models and metrics}

As mentioned above, all considered models consist of two parts: the regression part, which yields the fully-connected layers, and the physical part, in which differential equations are solved. The differentials equations include both, the parameters common to all stars $P^{C}$, and individual parameters $P^{I}$ that are different for each star.

The input data include the observed polar angles $\varphi_i:=\arctan\frac{\tilde{y}_i}{\tilde{x}_i}$ and physical polar angles $(\varphi_i)_{{\text{Phys}}}$,  within the limits in which we want to solve the regression problem.
For the Keplerian case the output of regression part is the predicted $\hat{\tilde{u}}_i$, which is compared with observed/target  $\tilde{u}_i:=\frac{1}{\tilde{r}_i}=\frac{1}{\sqrt{\tilde{x}_i^2+\tilde{y}_i^2}}$ using the MSE Loss function.

\begin{figure}[h!]
	\centering
	\begin{subfigure}{0.35\textwidth}
		\includegraphics[width=\linewidth]{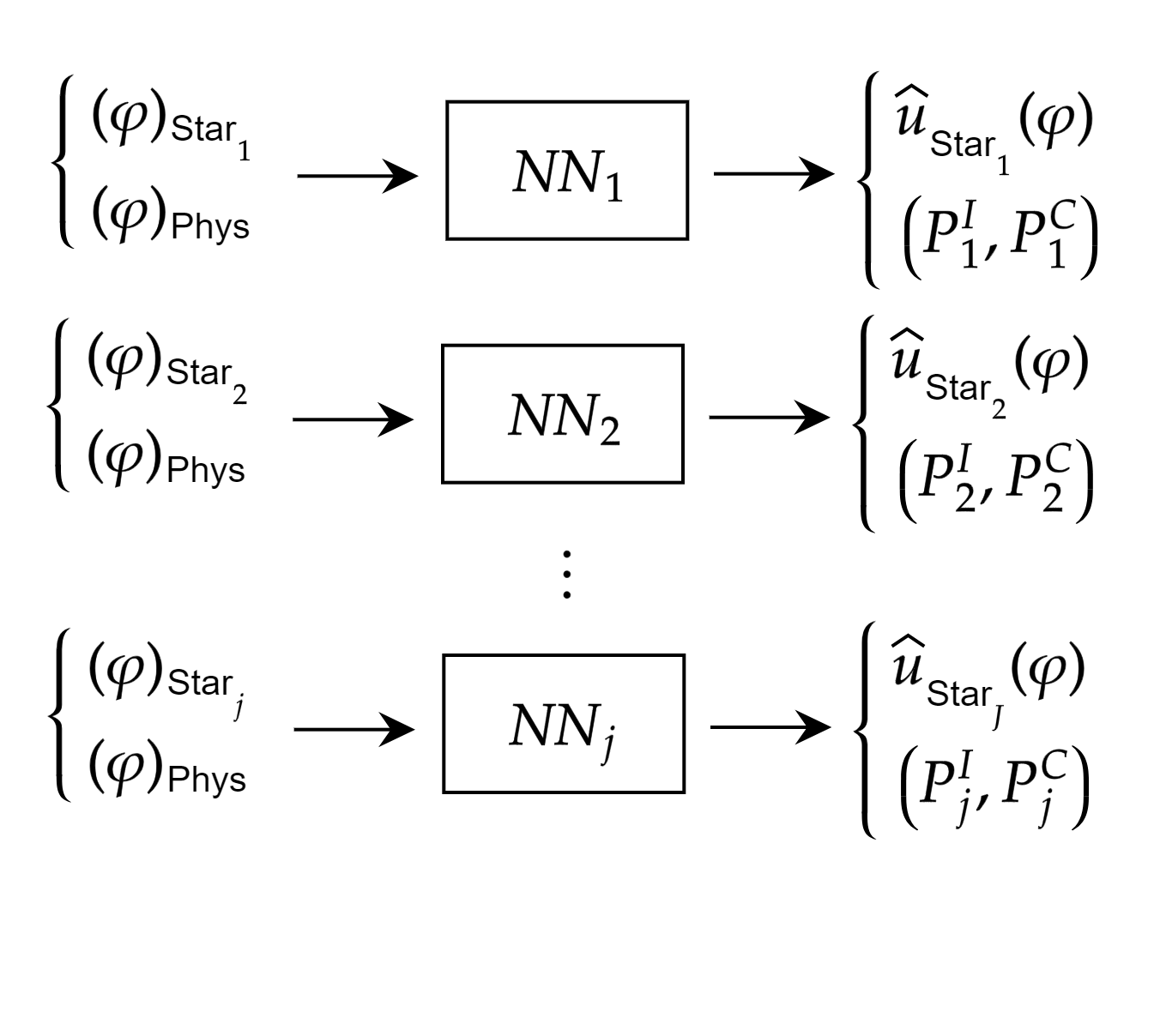}
		\caption{Individual training: Networks are trained separately for each star. They do not have any common parameters.}
	\end{subfigure}\hspace{0.1\textwidth}
	\begin{subfigure}{0.35\textwidth}
		\includegraphics[width=\linewidth]{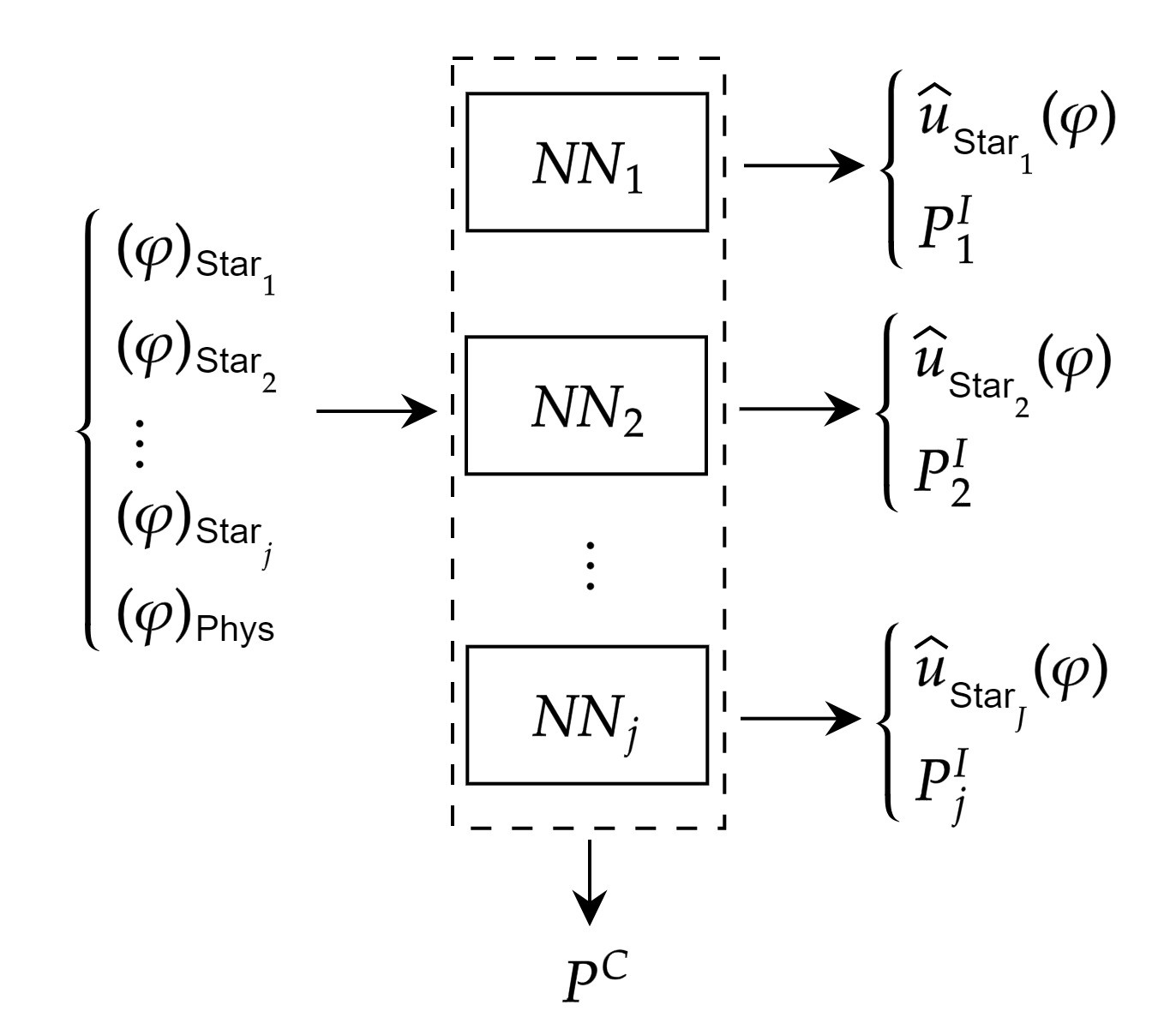}
		\caption{Parallel training: Networks are trained parallelly, so they have common parameters like the central mass.}
	\end{subfigure}
 \\
    \begin{subfigure}{0.8\textwidth}
		\includegraphics[width=\linewidth]{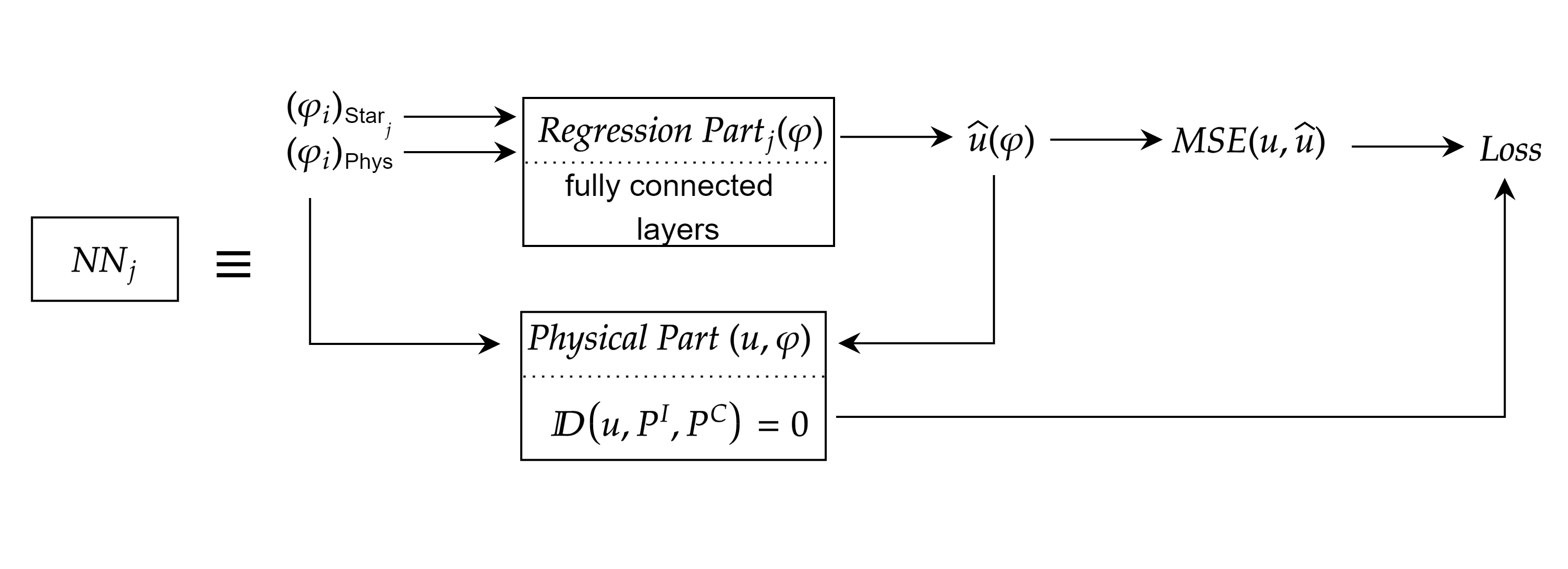}
		\caption{NN block: A single network which is used to train on a single star data. These blocks are used in both individual and parallel training cases.}
	\end{subfigure}
 \caption{Training scheme: $(\varphi)_{\text{Star}_j}$ are the polar angles of the $j$-th S-star, $(\varphi)_{\text{Phys}}$ is the physical polar angles set, i.e. the domain (wider than the observational domain) in which we want to predict the motion, $\varphi_i$ is the $i$-th polar angle for the corresponding star or physical angles, $P^{I}$ is the star individual parameters (such as eccentricity), $P^{C}$ is the common parameters (such as central body mass) and $\mathbb{D}(\cdot)$ is the physical model differential equations.}
 \label{fig:Schemes}
\end{figure}

And for the GR case, the Schwarzschild metric output was involved using the Darwin variable $\chi_i$ \cite{Chandra}. 
Two different training methods were used:\\
1) Parallel training, when the data of all stars are simultaneously fed into the input of copies of the neural network and in the physical part, the common training parameters $P^{C}$ (such as the mass of the central body) are the same parameter.\\
2) Individual training, the data of each star are involved separately and the training is sequential from star to star, in this case there are no common parameters in the physical part.\\
The schemes in Fig.(\ref{fig:Schemes}) illustrate how the training process works.

To obtain statistically significant results, the physical losses, taking into account the regularization coefficient, must be less than the values of the terms that enter into the losses. 

As for physical losses, we have used both the second and first order differential equations simultaneously (see~Eq.(\ref{eq:kepler}) or Eq.(\ref{eq:darwin_phi_chi})). The reason is that, the first order equations usually get stuck on a constant value after the extremum point, i.e. when $\frac{du}{d\varphi} = 0$ in Eq.(\ref{eq:kepler}) or $\frac{d\chi}{d\varphi} = 0$ in Eq.(\ref{eq:darwin_phi_chi}), because of the zero under the square root in those points. Using both equations has helped NN to avoid that problem and have a better performance. 

To estimate the performance of the models we use the metrics given by Eq.(\ref{eq:Metric})
\begin{equation}\label{eq:Metric}
\begin{split}
    &\mathcal{M}_{\text{model-data}}=\mathbb{E}\left[1-\frac{|u_{\text{Model}}-u_{\text{Star}}|}{u_{\text{Star}}}\right],\\
    &\mathcal{M}_{\text{data-physics}} = \mathbb{E}\left[1 - \frac{|u_{\text{Phys}}-u_{\text{Star}}|}{u_{\text{Star}}}\right],\\
    &\mathcal{M}_{\text{model-physics}} = \mathbb{E}\left[1 - \frac{|u_{\text{Model}}-u_{\text{Phys}}|}{\frac{1}{2}(u_{\text{Model}}+u_{\text{Phys}})}\right],\\
\end{split}
\end{equation}
where $u_{\text{Model}}$ is the predictions of the model, $u_{\text{Star}}$ is the used  data, and $u_{\text{Phys}} = \frac{1}{\hat{p}}(1+\hat{e}\cos(\varphi-\varphi_0))$ is the prediction by the obtained orbital parameters. These metrics, besides showing us how well the model has learned the data, also show how well the model has ``understood" the physics. Note that, while ``data" metrics are calculated for the given data points, the $\mathcal{M}_{\text{model-physics}}$ is calculated for the physical polar angles $(\varphi)_\text{Phys}$.

\section{Numerical experiments}

\subsection{Kepler case}

First, we consider the case of the Keplerian potential, 
\begin{equation}
    V(r)=-\frac{M}{r},
\end{equation}
i.e. we assume that the stars move along ellipses. In this case, the energy $E$ and the angular momentum $L$ are conserved, together with the normalized energy $\tilde{E} = \frac{E}{m}$ and angular momentum $\tilde{L} = \frac{L}{m}$ by the mass of the orbiting object $m$. 
\begin{equation}
\begin{split}
    &\tilde{E} = \frac{\dot{r}^2}{2} + \frac{r^2\dot{\varphi}^2}{2} - V(r), \\
    &\tilde{L} = r^2\dot{\varphi},
\end{split}
\end{equation}
where the dot is used for the time derivative \cite{Landau}. Substituting $u = \frac{1}{r}$, $e = \sqrt{1+\frac{2\tilde{E}\tilde{L}^2}{M^2}}$, and $p = \frac{\tilde{L}^2}{M}$ we arrive to the first equation in the Eq.(\ref{eq:kepler}), the second one is derived by differentiation of the first one \cite{Goldstein}. 
\begin{equation} \label{eq:kepler}
\begin{split}
    \left(\frac{du}{d\varphi}\right)^2 &= \frac{(e^2-1)}{p^2} - u^2 + \frac{2}{p}u, \\
    \frac{d^2u}{d\varphi^2} &= -u + \frac{1}{p}.
\end{split}
\end{equation}
In this case we carried out the individual training scheme with the training NN block shown in the Fig.(\ref{fig:Training-process}). Training parameters $P^{I}$ are the eccentricity and focal parameter. The network itself (regression part) consists of $4$ fully connected layers with $(32,64,32,1)$ nodes respectively and on all layers except the last one there is a $\tanh$ activation function. We considered the S1, S2, S9, S13 and S54 stars. The predicted orbital parameters $(\hat{e},\hat{p})$, together with estimated metrics Eq.(\ref{eq:Metric}) are shown in the Tab.(\ref{tab:KepPINN}).
\begin{figure}[h!]
	\centering
		\includegraphics[width=0.68\linewidth]{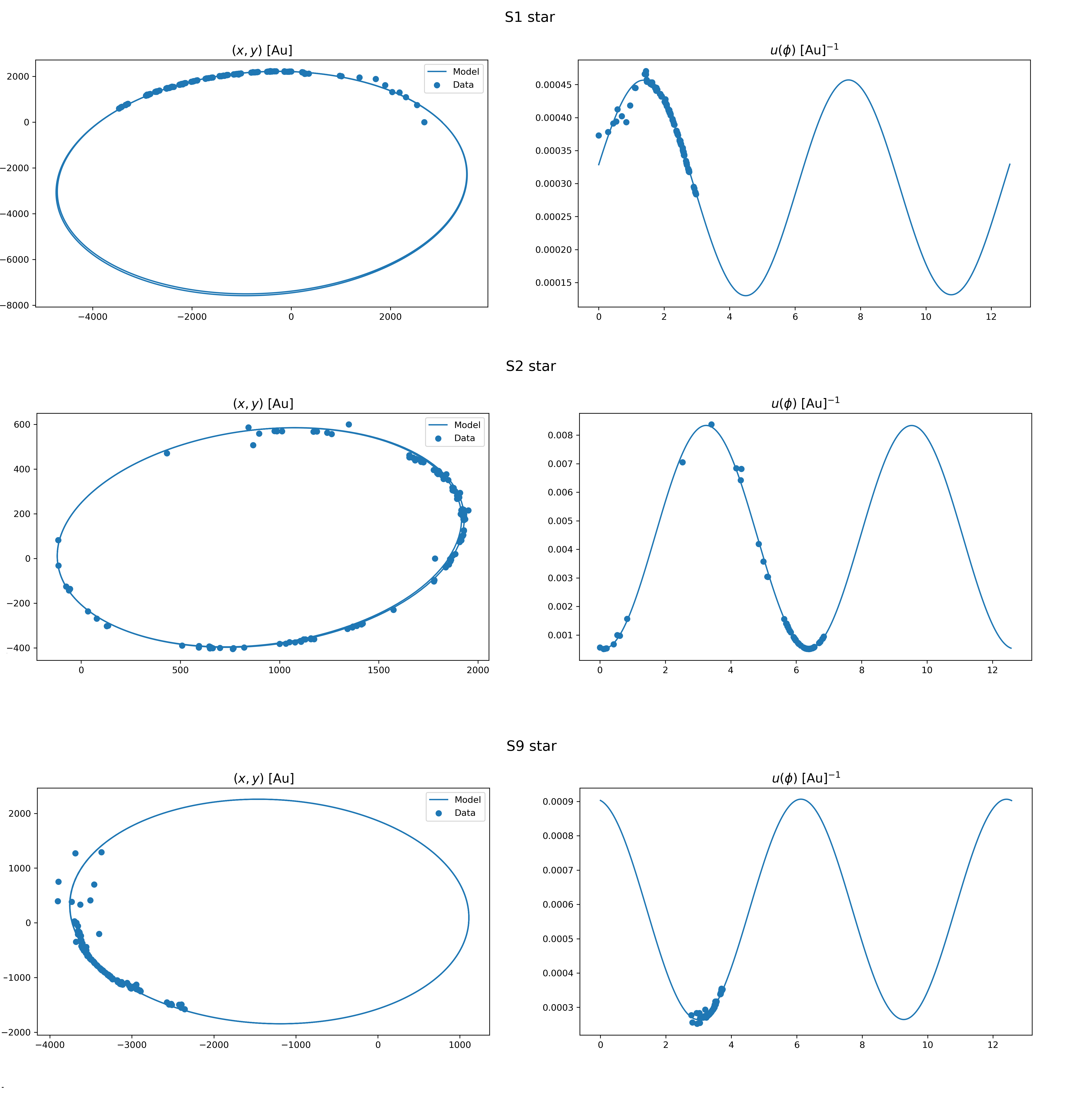}
		\includegraphics[width=0.68\linewidth]{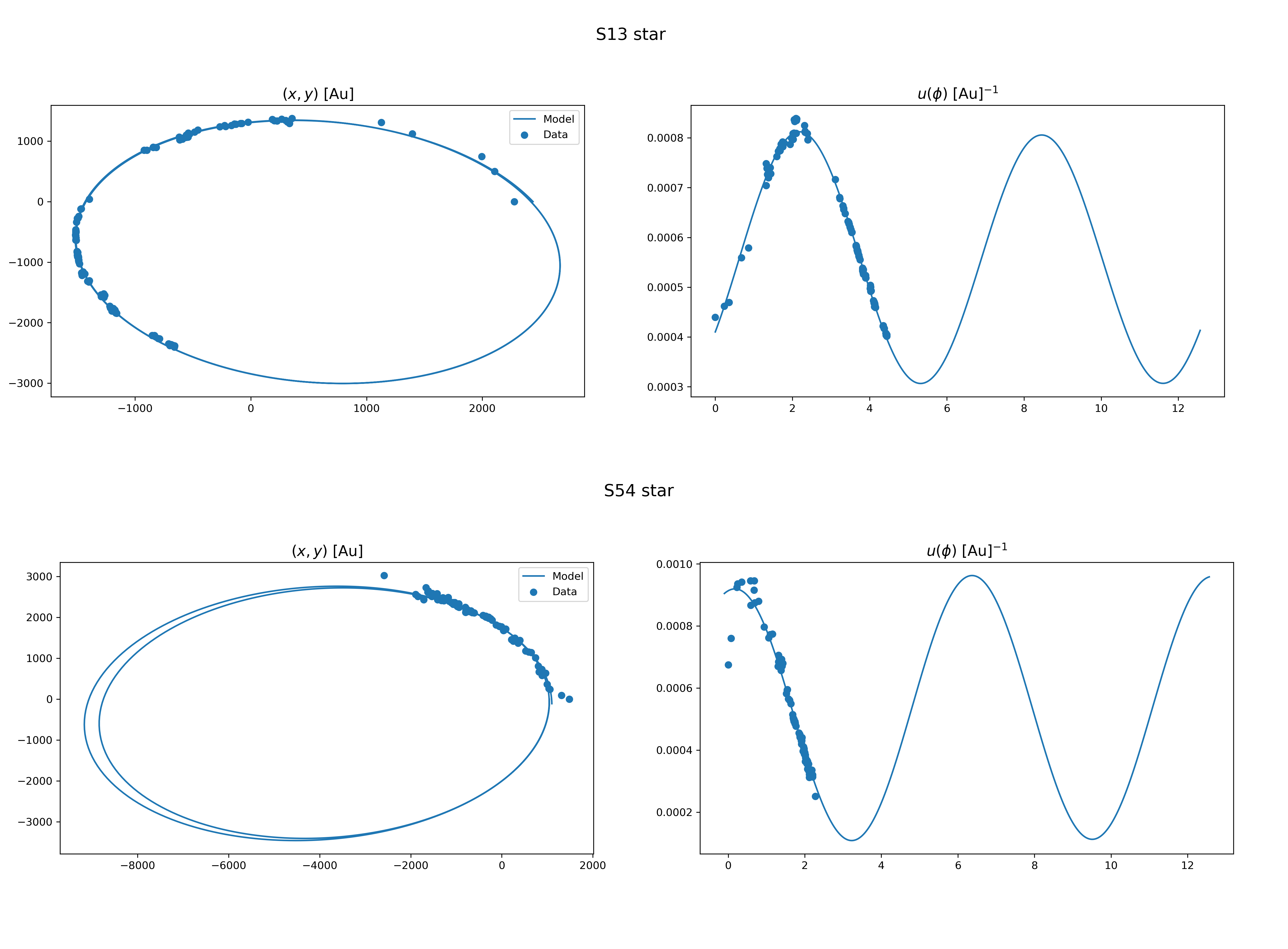}
 \caption{The model predictions for Kepler case: On the left, the predicted trajectories are shown in the Cartesian coordinates  $(x\,\,[\text{Au}],y\,\,[\text{Au}])$, where $(0, 0)$ is the central mass. On the right, the models' regression results are shown.}
 \label{fig:ParPredKep}
\end{figure}
\begin{table}[H]
    \centering
    \begin{tabular}{|c|c|c|c|c|c|c|c|}
        \hline
        Star Name & $e$ & $\hat{e}$ & $p\,\,[\text{Au}]$ & $\hat{p}\,\,[\text{Au}]$ & $\mathcal{M}_{\text{model-data}}$ &$\mathcal{M}_{\text{data-physics}}$&$\mathcal{M}_{\text{model-physics}}$\\
        \hline
        S1 & $0.556$  & $0.554$ & $3412$ & $3405$ & $0.9933$ & $0.9896$ & $0.9942$\\
        \hline
        S2 & $0.884$ & $0.872$ & $228$ & $226$ & $0.9871$ & $0.9763$ & $0.9894$\\
        \hline
        S9 & $0.644$ & $0.603$ & $1323$ & $1406$ & $0.9892$ & $0.9533$ & $0.9920$\\
        \hline 
        S13 & $0.425$ & $0.454$ & $1796$ & $1792$ & $0.9888$ & $0.9724$ & $0.9892$\\
        \hline
        S54 & $0.893$ & $0.712$ & $2017$ & $1862$ & $0.9662$ & $0.9359$ & $0.9738$\\
        \hline
    \end{tabular}
    \caption{Results for Kepler case: $(e,p)$ orbital parameters are from \cite{Gillessen}, $(\hat{e},\hat{p})$ the network prediction.}
    \label{tab:KepPINN}
\end{table}
The trajectories predicted by the model are shown in the Fig.(\ref{fig:ParPredKep}), which also shows the prediction of the models outside the observational domain.

\subsection{GR case}

\begin{figure}[h]
    \centering
    \begin{subfigure}{0.8\textwidth}
        \includegraphics[width=1\linewidth]{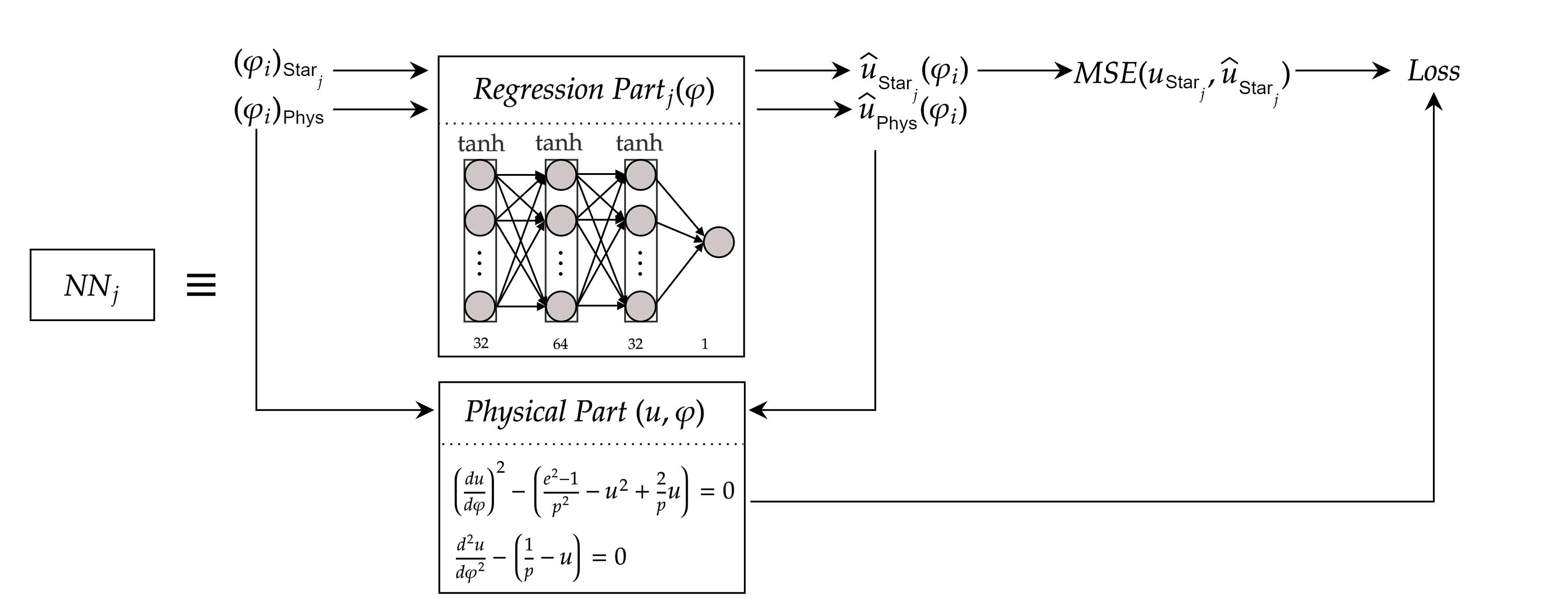}
        \caption{Training process for Kepler Case.}
    \end{subfigure}\\
     \begin{subfigure}{0.8\textwidth}
        \includegraphics[width=1\linewidth]{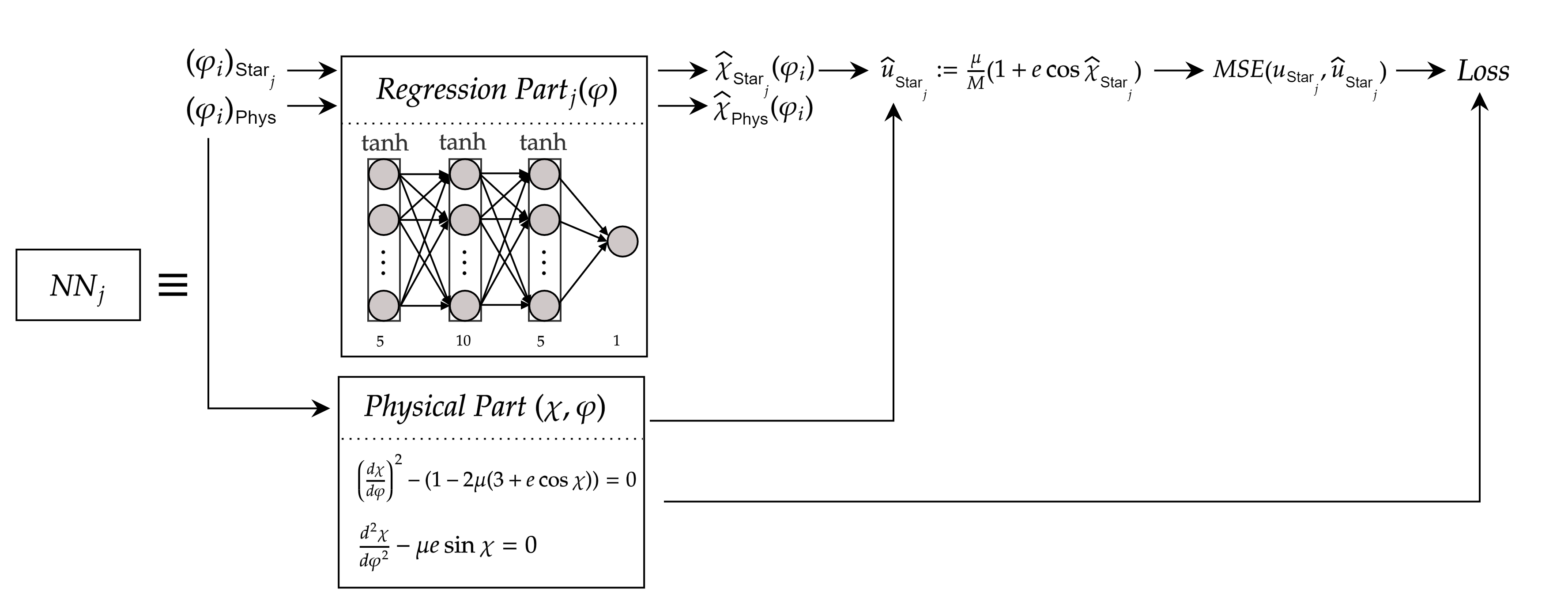}
        \caption{Training process for GR Case.}
    \end{subfigure}
    \caption{A more detailed NN block scheme for the Kepler and GR cases from the Fig.(\ref{fig:Schemes}(c)).}
    \label{fig:Training-process}
\end{figure}

The next step is to consider the Schwarzschild metric as a physical model for PINN. Following \cite{Chandra} and introducing the Darwin variable $\chi$, the equation of motion can be written as
\begin{equation} \label{eq:darwin_u_chi}
    u = \frac{\mu}{M}(1+e\cos\chi),
\end{equation} 
\begin{equation} \label{eq:darwin_phi_chi}
    \begin{split}
        \left(\frac{d\chi}{d\varphi}\right)^2 &= 1 - 2\mu(3+e\cos\chi), \\
        \frac{d^2\chi}{d\varphi^2} &= \mu e \sin{\chi},
    \end{split}
\end{equation}
where $\mu:=\frac{M}{p}$. In this case we carried out the parallel training scheme and the training NN block is shown in the Fig.(\ref{fig:Training-process}). 

The individual training parameters $P^{I}$ are the eccentricity and the focal parameter, and the common parameter $P^{C}$ is the mass of the central body. The network itself consists of $4$ fully connected layers with $(5,10,5,1)$ nodes respectively and on all layers except the last one, there is a $\tanh$ activation function. Moreover, artificial data is generated based on orbital parameters from \cite{Gillessen} for S13, S31, and S54 stars to close the orbit.

\begin{table}[h!]
    \centering
    \begin{tabular}{|c|c|c|c|c|c|c|c|}
        \hline
        Star Name & $e$ & $\hat{e}$ & $p\,\,[\text{Au}]$ & $\hat{p}\,\,[\text{Au}]$ & $\mathcal{M}_{\text{model-data}}$ &$\mathcal{M}_{\text{data-physics}}$&$\mathcal{M}_{\text{model-physics}}$\\
        \hline
        S2 & $0.884$ & $0.884$ & $228$ & $223$ & $0.9755$&$0.9768$&$0.9789$\\
        \hline
        S13 & $0.425$ & $0.418$ & $1796$ & $1706$ & $0.9844$&$0.9829$&$0.9974$\\
        \hline
        S31 & $0.550$ & $0.552$ & $2601$ & $2675$ & $0.9803$&$0.9826$ &$0.9835$\\
        \hline
        S54 & $0.893$ & $ 1.001 $ & $2017$ & $4.787$ & $0.9660$& $<0$ &$<0$\\
        \hline
    \end{tabular}
    \caption{Results for GR case: $(e,p)$ orbital parameters from \cite{Gillessen}, $(\hat{e},\hat{p})$ the network prediction.}
    \label{tab:GRPINN}
\end{table}
\begin{figure}[h!]
	\centering
	\begin{subfigure}{0.9\textwidth}
		\includegraphics[width=\linewidth]{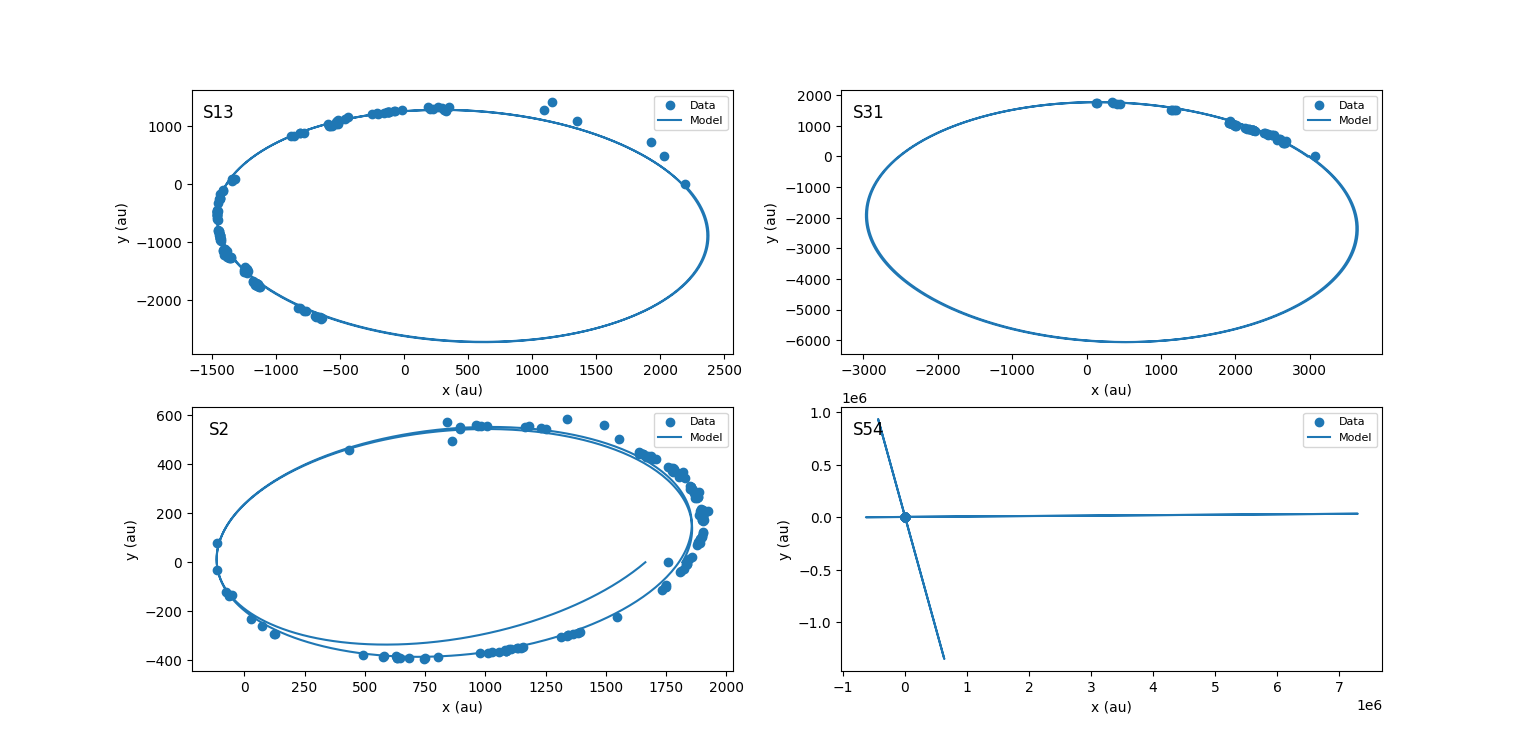}
		\caption{Predicted trajectories compared to the data as in the Fig. (\ref{fig:ParPredKep}).}
	\end{subfigure}\hspace{0.1\textwidth}
	\begin{subfigure}{0.9\textwidth}
		\includegraphics[width=\linewidth]{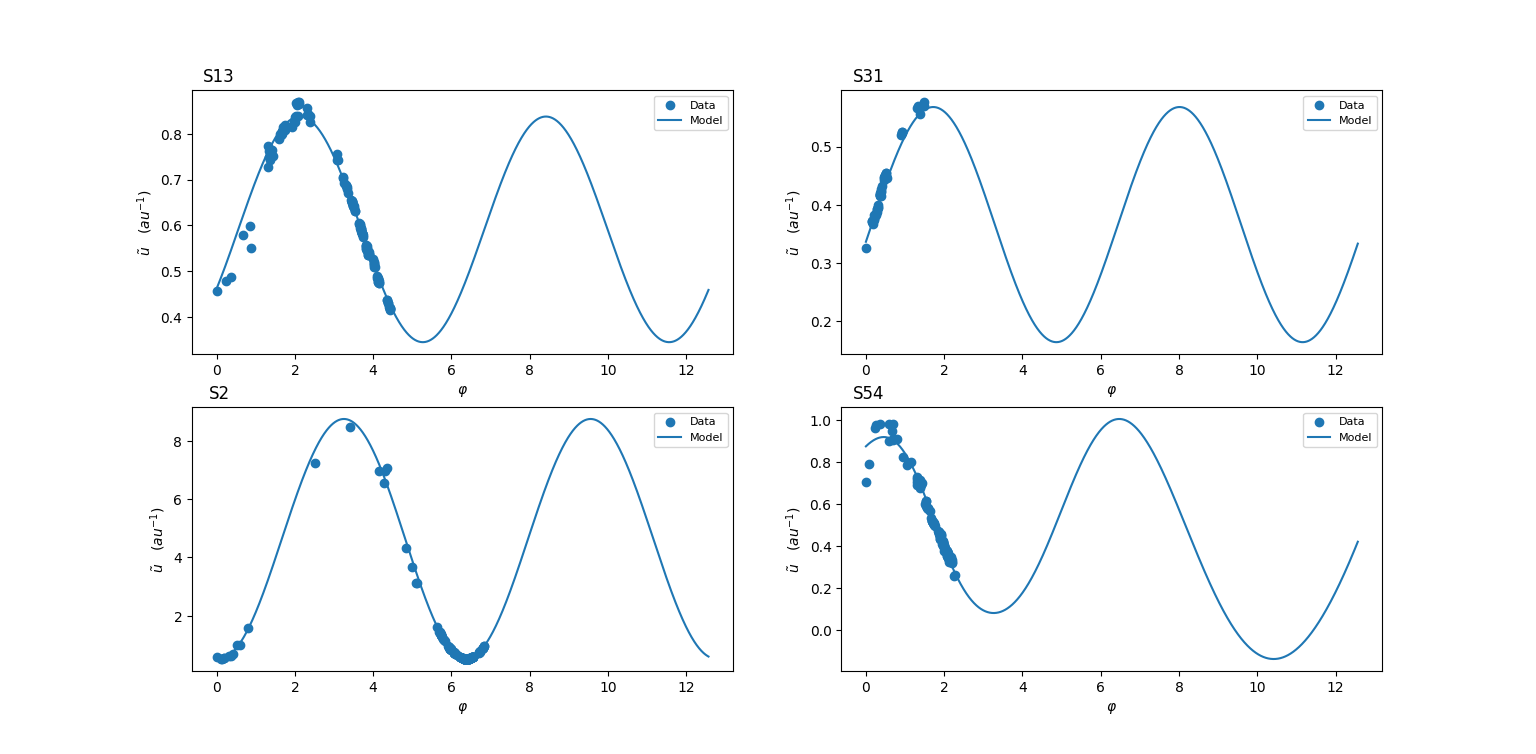}
		\caption{The regression results as in the Fig. (\ref{fig:ParPredKep}).}
	\end{subfigure}
 
 \caption{Model Predictions. \small{
 Note that the data domain used for the model's physical part training is set to be $[0, 4\pi]$, which means that besides learning the data, the model also extrapolates it as it was expected by PINN approach.}}
 \label{fig:ParPred}
\end{figure}
The results of model inference are shown in Fig. (\ref{fig:ParPred}). As for the star S54, we see that the model actually learned the data, but it failed to learn the physical part, thus extrapolating the trajectory to angles $>2\pi$. For the remaining stars, the model learned their orbital parameters well and was able to extrapolate outside the observed data.
\subsection{Precession \& $\Lambda$ constraint}
The S2 star data make it possible to check GR based on the precession \cite{GRAVITY}. The first order GR (Schwarzschild) correction, i.e. the following perturbation to the Keplerian potential Eq.(\ref{eq:SchPot}) is 

\begin{equation}\label{eq:SchPot}
    V(r)=-\frac{M}{r}+\frac{r_gL^2}{2}\frac{1}{r^3},
\end{equation}
where $L=\sqrt{pM}$ is the angular momentum of the test particle, corresponding to the Schwarzschild precession rate Eq.(\ref{eq:SchPrec}).

Statistical analysis of data within the first order PPN by the GRAVITY Collaboration \cite{GRAVITY}, reports a deviation from~Eq.(\ref{eq:SchPrec}) by a magnitude of $f_{\text{SP}}=1.10\pm 0.19$
\begin{align}\label{eq:f_SP_def}
    &\delta\varphi_{\text{GRAV}} = f_{\text{SP}}\cdot\delta\varphi_{\text{SP}}= 3f_{\text{SP}}\frac{\pi r_g}{a(1-e^2)};\\
    & f_{\text{SP}}=0,\quad \text{Keplerian case}, \nonumber\\
    & f_{\text{SP}}=1,\quad \text{GR case}.\nonumber
\end{align}

To find the contribution of the $\Lambda$ term to the precession, based on Eq.(\ref{eq:ModF}), we have the following perturbed potential
\begin{align}
    V(r)&=-\frac{M}{r}+\delta V_{\text{GR}}(r)+\delta V_{\Lambda}(r)\\
    &=-\frac{M}{r}+\frac{r_gL^2}{2}\frac{1}{r^3}-\frac{\Lambda}{6}r^2, \nonumber
\end{align}
which, along with $\delta\varphi_{\text{SP}}$, leads to an additional term $\delta\varphi_{\Lambda}$
\begin{equation} \label{eq:Lambda_prec}
    \delta\varphi_{\Lambda}=\frac{2a^3(1-e^2)^{\frac{1}{2}}\pi}{r_g}\Lambda.
\end{equation}
Based on the values reported by \cite{GRAVITY,Gillessen}, we find a constraint on the value of the cosmological constant as follows
\begin{equation}\label{eq:total_prec}
    \delta\varphi_{\text{SP}}+\delta\varphi_{\Lambda}=\delta\varphi_{\text{GRAV}},
\end{equation}
and the following upper constraint on the $\Lambda$  
\begin{equation}
    \Lambda\leq1.0\times10^{-36}\,\,[\text{m}]^{-2}.
\end{equation}

\begin{figure}[h!]
\centering
\includegraphics[width=\linewidth]{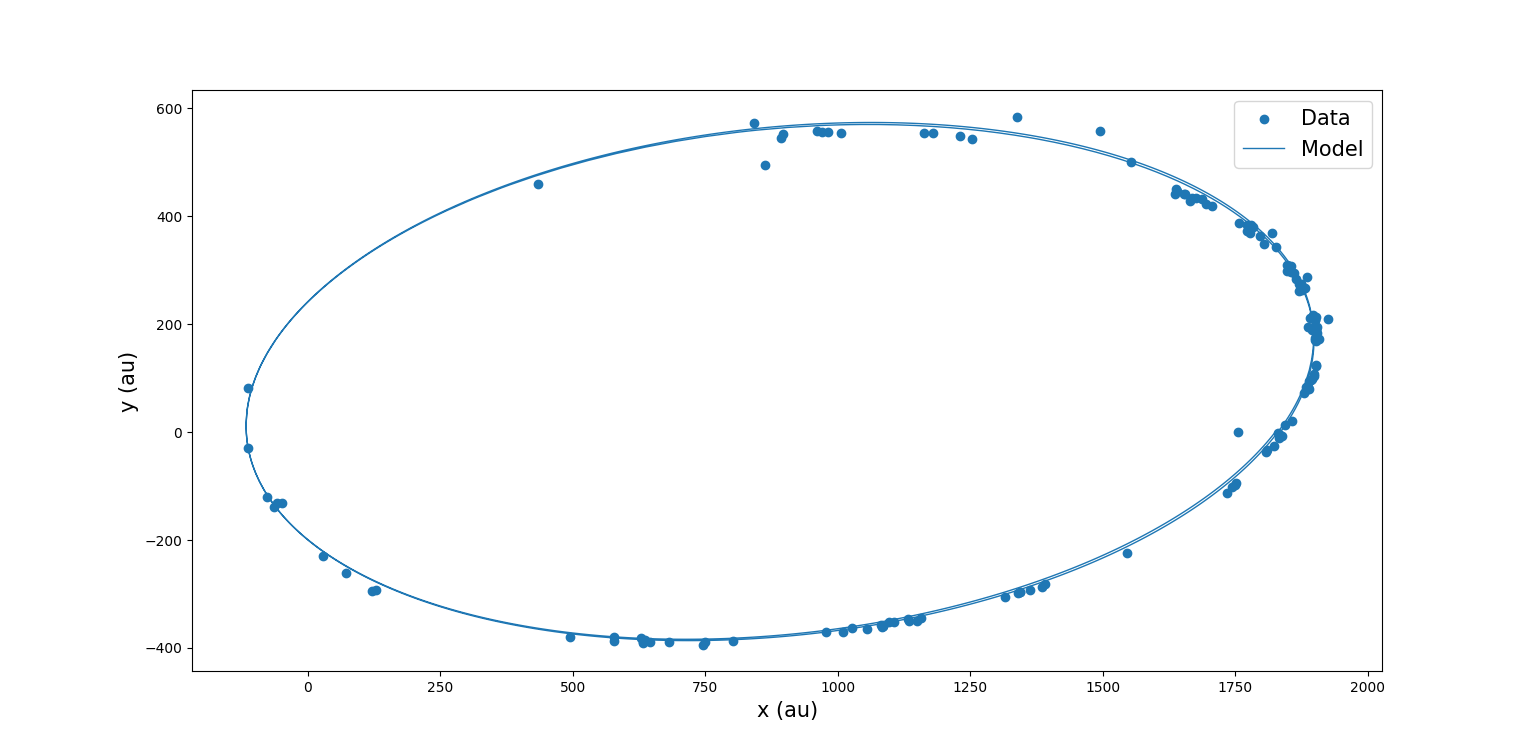}
\caption{The S2 star trajectory in  $(x\,\,[\text{Au}],y\,\,[\text{Au}])$ coordinates.}
\label{fig:S2_Traj}
\end{figure}

We can also find the constraint on $\Lambda$ from the PINN. After individual training on the S2 star, we get the following values for the parameters with their confidence intervals for the certain NN and the metrics
\begin{equation}
\begin{split}
    &\hat{e}= 0.88512\pm0.00001 ,\quad 
    \hat{p}=219.2\pm 0.2 \, [\text{Au}] ,\quad 
    \hat{M}= 0.04 \, [\text{Au}], \\
    &\mathcal{M}_{\text{model-data}} = 0.9865 ,\quad \mathcal{M}_{\text{data-phyics}} = 0.9881,\quad \mathcal{M}_{\text{model-physics}} = 0.9977. \\
\end{split}
\end{equation}
The predicted trajectory of motion of the S2 star is shown in the Fig. (\ref{fig:S2_Traj}).

It is important to note that, after a certain point during training, the value of $\hat{M}$ was fixed to be equal $0.04\,[\text{au}]$. Although the model was able to reach the given value and ``understand" the physical meaning of $\hat{M}$, which was indicated by the closeness ($\pm0.1'$) of precession rates calculated using physical parameters Eq.(\ref{eq:PhysPrec}) and regression Eq.(\ref{eq:RegPrec}) after that point (before the model had already obtained the values of eccentricity, focal parameter, and obtained the regression results, but the value of physical precession was an order of magnitude higher), its value was not stable due to the quality of the data. 

During training, we calculate the two values of the precession rate:
\begin{equation}\label{eq:RegPrec}
    \delta\varphi_{\text{Reg}}=\varphi\left(\min_1 \hat{u}\right)-\varphi\left(\min_0 \hat{u}\right)-2\pi,\quad \text{Precession rate of regression part}\,\, \hat{u}(\varphi).
\end{equation}
\begin{equation}\label{eq:PhysPrec}
    \delta\varphi_{\text{Phys}}=3\frac{\hat{r}_g}{\hat{p}}\pi,\quad \text{Precession rate of physical part}.
\end{equation}
Taking the moving average for every 500 epochs we obtain the following results
\begin{align}
    &\delta\varphi_{\text{Reg}}= 11.84';\quad\sigma_{\text{Reg}}=0.03',\\
    &\delta\varphi_{\text{Phys}}=11.82';\quad\sigma_{\text{Phys}}=0.02'.
\end{align}
During the calculations we took into account that for the significant results physical loss must be less than the individual terms that enter into it, and choose the step for $\varphi\in[0;4\pi]$ for which we calculate $\delta\varphi_{\text{Reg}}$ is so that, it is less than the difference $\delta\varphi_{\text{Reg}}-\delta\varphi_{\text{Phys}}$. Using the equations Eq.(\ref{eq:Lambda_prec}) and Eq.(\ref{eq:total_prec}), also taking $\delta\varphi_{\text{Reg}}(+3\sigma_{\text{Reg}})$ as the total precession rate and $\delta\varphi_{\text{Phys}}(-3\sigma_{\text{Phys}})$ as the Schwarzschild
precession rate, we obtain for the $\Lambda$ the following upper constraint
\begin{equation}
    \Lambda\leq5.8\times 10^{-38}\,\,[\text{m}]^{-2}.
\end{equation}

\section{Conclusions}

We studied the dynamics of the S-stars using neural networks PINN, aiming first to reveal the values of the orbital parameters of each star. Both Keplerian and General Relativity dynamics were considered to reveal their differences for the given network architecture. It was shown that with a given physical model, with good accuracy, one can obtain the orbital parameters of stars, as well as a regression problem can be solved providing the dependence of $u(\varphi)$.

The neural network was able to ``see" the Schwarzschild precession for S2 star, which made it possible to find the precession rate for both, based on the regression part and the physical part. And since the regressed part is more ``flexible" and is directly related to the observational data, the difference in the values of the precession rates can be attributed to an additional precession that occurs due to terms not entered in the physical model. Specifically, as such a contribution we considered the gravity modification with the cosmological constant $\Lambda$ in Eq.(2). Ultimately, this procedure enables us to find a constraint on the cosmological constant for the current data accuracy, being in agreement with its adopted value.
The same analysis was carried out using the results for the precession obtained by the GRAVITY collaboration \cite{GRAVITY}.

Our analysis reveals the efficiency of the neural networks in the study of the S-star dynamics and that stronger constraints on GR and gravity modifications can be expected from forthcoming observational data.

\section{Acknowledgment}
Sh.K. is acknowledging the ANSEF grant 23AN:PS-astroth-2922.

\section{Data Availability Statement} 
Data sharing not applicable to this article as no datasets were generated or analysed during the current study.

\newpage
\end{document}